\documentclass[reprint,prl,floatfix,aps]{revtex4-1}

\usepackage{graphicx}
\usepackage{epstopdf} 

\usepackage{url,lineno}
\usepackage{dcolumn}
\usepackage{bm}
\usepackage{tikz}
\newcommand*\circled[1]{\tikz[baseline=(char.base)]{
            \node[shape=circle,draw,inner sep=2pt] (char) {#1};}}

\begin{document}

\title{PolyPIC: the Polymorphic-Particle-in-Cell Method for Fluid-Kinetic Coupling}

\author{Stefano~Markidis$^{1}$, Vyacheslav~Olshevsky$^{1}$, Chaitanya~Prasad~Sishtla$^{1}$, Steven~Wei-der~Chien$^{1}$,  Erwin~Laure$^{1}$ and Giovanni~Lapenta$^{2}$}
\affiliation{$^{1}$Computational Science and Technology Department, Electrical Engineering and Computer Science School, KTH Royal Institute of Technology, Stockholm, Sweden \\
$^{2}$Centre for Plasma Astrophysics, KU Leuven, Leuven, Belgium}

\date{\today}

\begin{abstract}
Particle-in-Cell (PIC) methods are widely used computational tools for fluid and kinetic plasma modeling. While both the fluid and kinetic PIC approaches have been successfully used to target either kinetic or fluid simulations, little was done to combine fluid and kinetic particles under the same PIC framework. This work addresses this issue by proposing a new PIC method, PolyPIC, that uses polymorphic computational particles. In this numerical scheme, particles can be either kinetic or fluid, and fluid particles can become kinetic when necessary, e.g. particles undergoing a strong acceleration. We design and implement the PolyPIC method, and test it against the Landau damping of Langmuir and ion acoustic waves, two stream instability and sheath formation. We unify the fluid and kinetic PIC methods under one common framework comprising both fluid and kinetic particles, providing a tool for adaptive fluid-kinetic coupling in plasma simulations.
\end{abstract}

\maketitle

\section{Introduction}
Particle-in-Cell (PIC) methods are among the most popular computational methods for plasma simulations. There are two major families of PIC methods: the first one comprises the fluid PIC methods that solve the plasma equations in fluid approximation such as magnetohydrodynamic (MHD); the second one includes kinetic PIC methods for solving the kinetic equations of collisionless plasmas. 

Quite surprisingly, the fluid and kinetic PIC methods originated and evolved rather independently. The fluid PIC method was first developed in the Sixties by Harlow to solve the fluid equations by advecting fluid quantities (mass, momentum and energy) with computational particles~\citep{harlow1962particle}. New fluid PIC schemes were developed to decrease numerical diffusion and to solve plasma and material science problems \citep{Brackbill:1991,Bacchini:2017}. In particular, the fluid PIC method eventually merged with the Material Point Method for simulating continuous materials~\citep{sulsky1995application}. FLIP MHD~\citep{brackbill1991flip} and Slurm~\citep{olshevsky2018slurm} are among the most successful fluid PIC codes for plasma simulations.

The first kinetic PIC methods were developed by Buneman and Dawson in the late Fifties and early Sixties to model collisionless plasma~\citep{buneman1959dissipation, dawson1962one}. After their inception, the development of kinetic PIC methods focused more on increasing numerical stability with large simulation time steps~\citep{brackbill1982implicit,drouin2010particle} and ensuring energy conservation~\citep{markidis2011energy,chen2011energy,lapenta2011particle}. VPIC~\citep{bowers20080} and iPIC3D~\citep{markidis2010multi,peng2015formation,peng2015energetic} are among the most widely used kinetic PIC codes for plasma simulations.

Although the two PIC families developed independently with little cross-fertilization, they share the same conceptual framework~\citep{brackbill2005particle} as they both use computational particles for solving the advection term in the governing equations. The goal of this work is to unify and couple the fluid and kinetic PIC methods under the same framework by allowing the PIC computational particles to be polymorphic and have either fluid or kinetic nature. A major result of this work is the possibility of a fluid particle to become kinetic enabling a seamless fluid-kinetic coupling within the PIC method.

How to couple fluid and kinetic models within the same computational framework is a topic of several recent studies and projects~\citep{henri2013nonlinear,lapenta2013swiff,jordanova2017specification,innocenti2017progress}. The fluid-kinetic PIC method is an extension of the implicit-moment PIC method~\citep{brackbill1982implicit} using particles to calculate the pressure tensor without relying on an ad-hoc equation of state~\citep{markidis2014fluid}. The MHD-EPIC (MHD with Embedded PIC) by~\citet{daldorff2014two} is probably the most successful realization of coupling a PIC code, iPIC3D, and a fluid code, BATS-R-US, under the Space Weather Modeling Framework (SWMF)~\citep{toth2005space}. In this implementation, the whole computational domain is modeled by solving the MHD equations while the selected regions of space where kinetic effects are important are modeled with the kinetic PIC method. The coupling is achieved by feeding the MHD results to the kinetic solver via boundary conditions while the kinetic results replace the MHD in the specified domain. The MHD-EPIC method has been successfully used to model planetary magnetospheres~\citep{Toth:etal:2016, chen2017global, toth2017scaling,ma2018reconnection}. 

This work is inspired by the Vlasov spectral methods using Hermite polynomials~\citep{delzanno2015multi} and combining fluid and kinetic models within the same framework~\citep{vencels2016spectralplasmasolver}. In fact, by dynamically changing the number of Hermite polynomials during the simulation, it is possible to smoothly transition from fluid to kinetic within the same framework~\citep{vencels2015spectral}. In the spirit of these Vlasov spectral methods, this work investigates how to smoothly transition from fluid to kinetic within the PIC framework by transforming fluid particles to kinetic particles.

We propose a novel PIC method, PolyPIC, using polymorphic computational particles that allow for a smooth transition from fluid to kinetic approach. The PolyPIC method is tested against four standard benchmark problems, showing that it provides a seamless transition from fluid to kinetic modeling under the same computational framework. The paper is organized as follows. Section~\ref{method} presents the PolyPIC governing equations, explains the algorithm, discretization and implementation. Section~\ref{results} presents the results of testing the PolyPIC method against standard benchmark problems. Finally, Section~\ref{conclusion} summarizes this work, discusses its potential and limitations, and outlines future developments.

\section{The Polymorphic-Particle-in-Cell Method}
\label{method}
In this section, we present the governing equations, the Polymorphic-Particle-in-Cell (PolyPIC) algorithm, its discretization and numerical stability conditions.

\subsection{Governing Equations}
The microscopic state of a plasma species $\alpha$ (electrons or ions) is described by the distribution function $f_\alpha(\mathbf{x},\mathbf{v},t)$ that provides the number of plasma particles in the neighborhood of the position $\mathbf{x}$ and velocity $\mathbf{v}$ in the six-dimensional coordinate-velocity phase space. The evolution of a collisionless plasma species $\alpha$ with mass $m_\alpha$ and charge $q_\alpha$ in the presence of an electric field $\mathbf{E}$ (for sake of simplicity, magnetic field is absent) is governed by the Vlasov equation, which is a conservation law for the phase space density:
\begin{equation}
\label{Vlasov_kin}
\frac{\partial f_\alpha(\mathbf{x},\mathbf{v},t)}{\partial t} + \mathbf{v} \cdot \nabla_{\mathbf{x}} f_\alpha(\mathbf{x},\mathbf{v},t)+  \frac{q_\alpha}{m_\alpha}  \mathbf{E} \cdot  \nabla_{\mathbf{v}} f_\alpha(\mathbf{x},\mathbf{v},t) = 0,
\end{equation}
where $t$ is time, $\mathbf{x}$ and $\mathbf{v}$ are the coordinates in the position and velocity spaces. 
The Vlasov equation provides the full time-dependent description of the plasma and allows modeling of all plasma processes which depend on particle velocity, such as resonance phenomena. However, the numerical solution of the Vlasov equation in multi-dimensional phase space is often prohibitive as it requires to resolve the smallest time and space scales in the system.

The macroscopic state of the plasma species $\alpha$ can be conveniently characterized by the fluid approach, in terms of its mass density ($\rho_{\alpha,m}$) or charge density ($\rho_{\alpha,c}$), fluid bulk velocity ($\mathbf{u}_\alpha$), internal energy ($I_\alpha$), and pressure ($p_\alpha$). If we neglect heat flux, heat sources, viscosity and external forces, the evolution of such system is determined by the conservation of mass, momentum, and energy, accompanied with an equation of state (EoS):
\begin{equation}
\label{eq:fluid}
\begin{array}{l}
\displaystyle{\frac{\partial \rho_{\alpha,m} (\mathbf{x},t)}{\partial t }} + \left(\mathbf{u}_\alpha (\mathbf{x},t) \cdot \nabla \right)\rho_{\alpha,m} (\mathbf{x},t) \\ = - \rho_{\alpha,m} (\mathbf{x},t)\nabla \cdot \mathbf{u}_\alpha (\mathbf{x},t) \\[12pt]
\rho_{\alpha,m} (\mathbf{x},t) \left[\displaystyle{\frac{\partial \mathbf{u}_\alpha(\mathbf{x},t)}  {\partial t}} + \left(\mathbf{u}_\alpha(\mathbf{x},t) \cdot \nabla \right) \mathbf{u}_\alpha(\mathbf{x},t) \right] \\ =  - \nabla p_\alpha(\mathbf{x},t) + \rho_{\alpha,c} \mathbf{E} \\[12pt]
\rho_{\alpha,m}(\mathbf{x},t) \left[\displaystyle{\frac{\partial I_\alpha(\mathbf{x},t)} {\partial t}} +  \left( \mathbf{u}_\alpha \cdot \nabla \right) I_\alpha(\mathbf{x},t) \right] \\ =  - p_\alpha \nabla \cdot \mathbf{u}_\alpha \\[12pt]
p_\alpha(\mathbf{x},t) = p_\alpha(\rho_{\alpha,m},I_\alpha).
\end{array}
\end{equation}
The above fluid quantities used to describe the plasma are essentially the averages of the distribution function $f_\alpha(\mathbf{x},\mathbf{v},t)$ in the velocity space. The fluid equations can be derived from the first three moments of Vlasov equation (see, e.g., \citet{Freidberg:1987}). To compute each moment, the Vlasov equation is multiplied by the corresponding power of $\mathbf{v}$ and integrated over the velocity space:
\begin{equation}
\begin{array}{l}
\{ \rho_{\alpha,m},  \rho_{\alpha,c}, {\bf u}_\alpha, p_\alpha \}   =  \int \{m_\alpha, q_\alpha, \mathbf{v}, m_\alpha (\mathbf{v} - {\bf u}_\alpha)(\mathbf{v} - {\bf u}_\alpha)  \}  \\[8pt] f_\alpha(\mathbf{x},\mathbf{v},t) d{\bf v}.
\end{array}
\end{equation}
The fluid approximation drastically simplifies the treatment of the evolution of plasma, but loses information about individual particles, therefore it can not describe microscopic phenomena.

In both kinetic and fluid electrostatic models, the electric field $\mathbf{E}$ can be described in terms of the electrostatic potential $\Phi$
\begin{equation}
\label{eq:electric}\mathbf{E}(\mathbf{x},t)=-\nabla\Phi,
\end{equation}
and is governed by the Poisson's equation
\begin{equation}
\label{eq:poisson}\nabla^2 \Phi = - \rho_{net}(\mathbf{x},t),
\end{equation}
where $\rho_{net}(\mathbf{x},t) = \sum_\alpha \rho_{\alpha,c}(\mathbf{x},t)$ is the net charge density of the plasma.

\subsection{Algorithm}
The algorithms of fluid PIC methods are discussed in detail in Refs.~\citep{brackbill1988flip, Bacchini:2017,olshevsky2018slurm}, while kinetic PIC methods are extensively presented in two textbooks~\citep{birdsall2004plasma,hockney1988computer}. In essence, both fluid and kinetic PIC methods are semi-Lagrangian numerical methods. The main idea of the PIC method is in using computational particles to calculate the advection term ($\mathbf{u} \cdot \nabla$) of the governing equation(s), the Vlasov equation~\ref{Vlasov_kin}, or the fluid equations~\ref{eq:fluid}. This step for calculating the advection term is called \emph{Lagrangian} step. The remaining terms of governing equations are solved on a discrete computational grid. This other step is called \emph{Eulerian} step. 

At each PIC computational cycle, the advection is computed by updating computational particle positions and velocities $\mathbf{x}_p$, $\mathbf{v}_p$ for both fluid and kinetic particles. In addition, particle internal energy $e_p$ is also updated in the case of fluid particles. Each polymorphic particle carries mass $m_p$ and charge $q_p$. These two quantities are calculated initially dividing the total charge and mass per cell by the number of particles per cell. While it is possible to have different $m_p$ and $q_p$, particles belonging to the same species have the same $m_p$ and $q_p$ in this work.

In the fluid PIC, the mass density ($\rho_m$), fluid velocity (${\bf u}$), internal energy ($I$), and pressure ($p$) are defined on the grid. On the other hand, only the charge density ($\rho_c$) is defined in the electrostatic kinetic PIC method. Because PolyPIC combines fluid and kinetic PIC methods, $\rho_m$, $\rho_c$, ${\bf u}$, $I$, and $p$ are defined on the grid in PolyPIC. In addition, the electrostatic field quantities, electric field (${\bf E}$) and electrostatic potential (${\Phi}$), are defined on the grid.

\begin{figure}[h!]
\begin{center}
\includegraphics[width=\columnwidth]{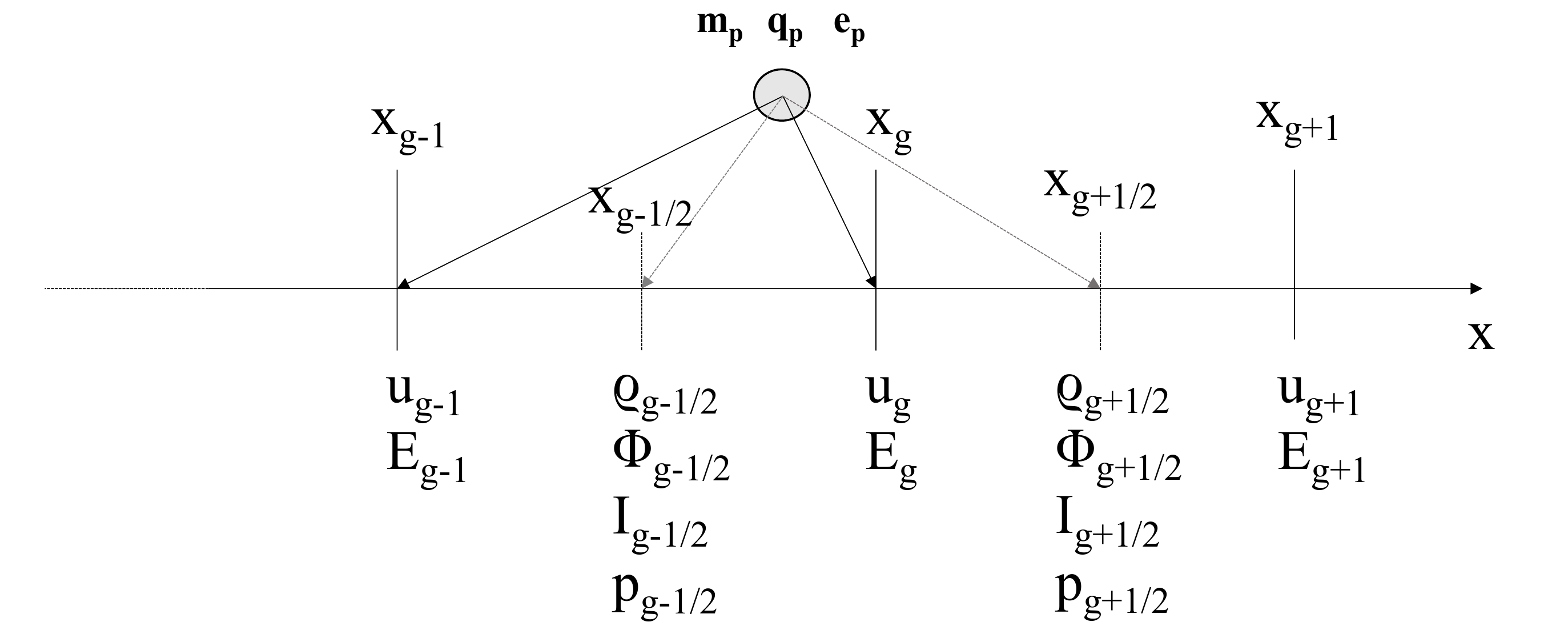}
\end{center}
\caption{Spatial discretization of PolyPIC method.}\label{fig:discretization}
\end{figure}

Our algorithm uses a staggered grid where $\mathbf{u}$ and $\mathbf{E}$ are defined on the grid nodes with subscripts $...,g-1, g, g+1,...$, while $\rho_c$, $\rho_m$, $\Phi$, and $I$ are defined on the centers of grid cells with subscripts $...,g-1/2, g+1/2,...$, as illustrated in Figure~\ref{fig:discretization}. Such discretization makes computation of gradients straightforward (the derivative of a cell-based quantity is a node quantity, and vice versa). In addition, a staggered grid is necessary to keep the magnetic field solenoidal without using artificial divergence cleaning, when the algorithm is extended to magnetized plasmas.

At any time in the PIC algorithm, it is possible to move from particle quantities to grid quantities simply using interpolation functions. Properties on grid points $\mathbf{x}_g$ are calculated by means of the interpolation functions $W({\bf x}_g-{\bf x}_p)$ (dropping the $\alpha$ subscript in the notation)
\begin{equation}
\label{eq:interp}
\{ \rho_m,  \rho_c, {\bf u}, I \}_g  =  \sum_p^{N_p} \{m_p, q_p, \mathbf{v}_p, e_p \} W({\bf x}_g-{\bf x}_p).
\end{equation}

Several interpolation functions can be used. In this work, piece-wise linear interpolation functions~\citep{birdsall2004plasma,hockney1988computer} are used:
\begin{equation}
 \label{eq:interpW}
W({\bf x}_g-{\bf x}_p) =
\left\{
\begin{array}{l}
 1 - |{\bf x}_g-{\bf x}_p|/\Delta x \quad \textup{if} \quad |{\bf x}_g-{\bf x}_p| < \Delta x \\
0  \quad \textup{otherwise}.
\end{array}
\right.
\end{equation}

In the fluid PIC method, the pressure on each grid point is derived from $I$ using and Equation of State (EoS). In this work, we use the ideal gas EoS:
\begin{equation}
p = \rho_m I (\gamma - 1)
\end{equation}
where $\gamma = c_p/c_V$ is the specific heats ratio.

\begin{figure*}[htp]
\begin{center}
\includegraphics[width=\textwidth]{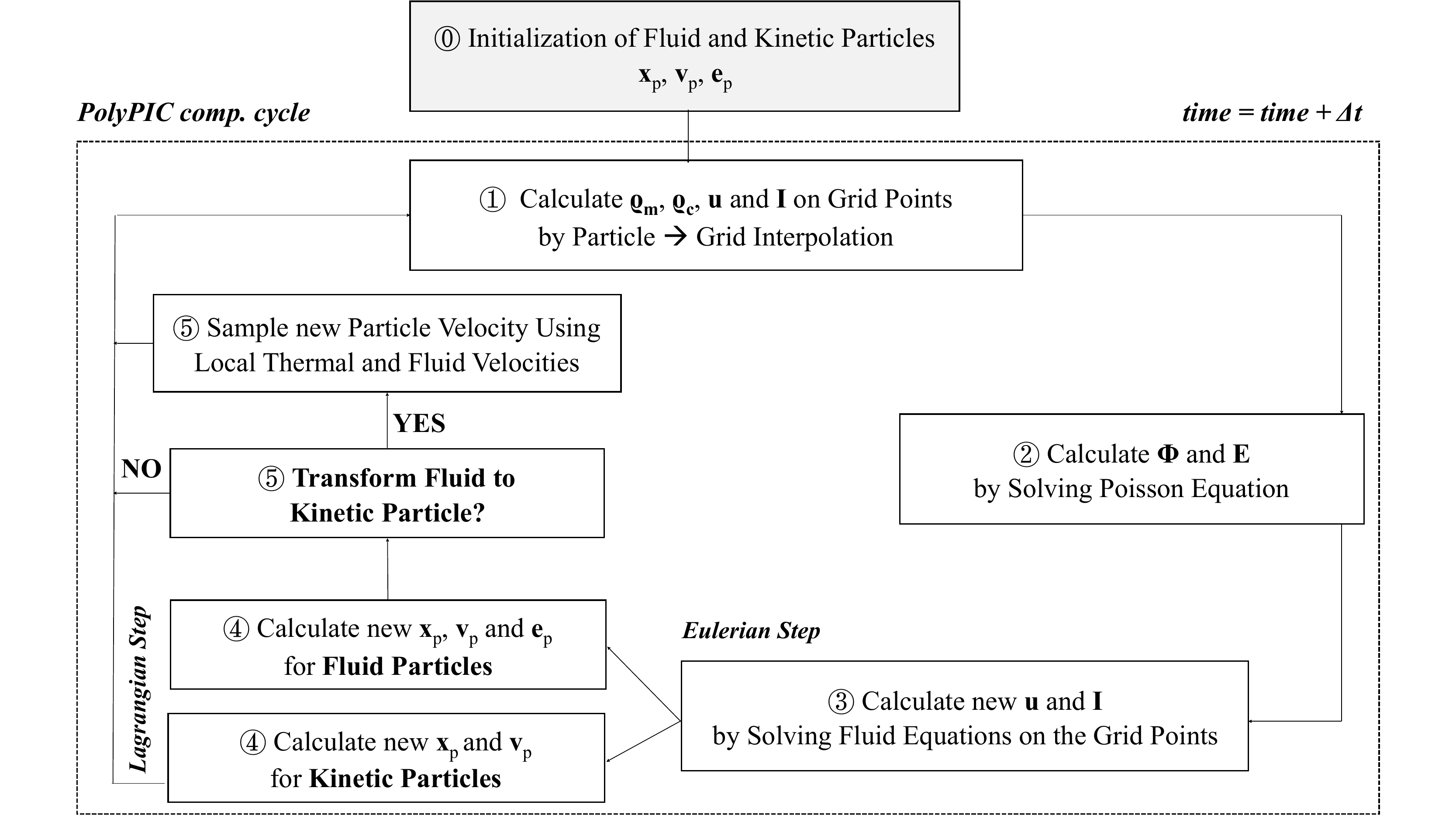}
\end{center}
\caption{The computational cycle of the PolyPIC method.}\label{fig:Algo}
\end{figure*}

The PolyPIC method comprises an initialization (\circled{0} in Figure \ref{fig:Algo}) for setting up the simulation parameters and a computational cycle that is repeated at each simulation time step. The computational cycle consists of five stages \circled{1}-\circled{5}, as illustrated in Figure~\ref{fig:Algo}.


{\bf \circled{0} Initialization.} During the initialization phase, fluid quantities (densities and fluid velocity) are defined on the grid and particles are set to be either fluid or kinetic. Particle positions are typically initialized as uniform in space. If particle is fluid, its mass and charge are determined from the local mass and charge densities, while its velocity is set to the local fluid velocity ${\bf u}$. If particle is kinetic, its charge and mass are still calculated from local densities, but its velocity is randomly sampled from a Maxwellian distribution centered at ${\bf u}$, with the variance equal to the thermal velocity. 

After the initialization, the following five phases are carried out at each computational cycle. 

{\bf \circled{1} Interpolation Particles $\rightarrow$ Grid.} The values of the fluid quantities on the grid points ($\rho_m$, $\rho_c$, ${\bf u}$ and $I$) are computed using the particle to grid interpolation functions (Equations \ref{eq:interp} and Figure~\ref{fig:discretization}). It is important to note that both kinetic and fluid particles participate in this interpolation step. Because of the thermal spread of kinetic particles, the quantities interpolated from kinetic particles are affected by thermal noise. 

{\bf \circled{2} Electric Field Calculation on the Grid.} After the particle to grid interpolation, it is possible to calculate the net charge density $\rho_{net} = \sum_\alpha \rho_{c,\alpha}$. The electrostatic potential $\Phi$ is computed by solving the Poisson's equation (Equation~\ref{eq:poisson}) on the grid. In this work, we use one dimensional geometry and finite difference discretization of Equation~\ref{eq:poisson} resulting in an algebraic equation for each grid point $g+1/2$:
\begin{equation}
\frac{\Phi_{g-1/2} - 2 \Phi_{g+1/2}  + \Phi_{g+3/2}} {\Delta x ^ 2}  = - \rho_{net, g+1/2}.
\end{equation}
The set of equations for each grid point constitutes a tridiagonal linear system that can be solved to find $\Phi$. After the solution of the linear solver, the electric field is calculated from $\Phi$ by discretizing Equation~\ref{eq:electric}) with finite difference:
\begin{equation}
E_g = - \frac{\Phi_{g+1/2} - \Phi_{g-1/2}} {\Delta x}. 
\end{equation}

{\bf \circled{3} Update Grid Quantities (Eulerian Step).} In this phase, the new ${\bf u}$ and $I$ values are calculated solving the fluid equations on the grid without the advection term. The momentum and energy fluid equations to be solved on the grid are:
\begin{equation}
\label{eq:fluidUpdate}
\begin{array}{l}
\rho_{m} (\mathbf{x},t) \displaystyle{\frac{\partial \mathbf{u}(\mathbf{x},t)}  {\partial t}} = - \nabla ( p (\mathbf{x},t)  + \mu(\mathbf{x},t) )+ \rho_{c}(\mathbf{x},t) \mathbf{E}(\mathbf{x},t) \\[12pt]
\rho_{m}(\mathbf{x},t) \displaystyle{\frac{\partial I(\mathbf{x},t)} {\partial t}} =  - (p(\mathbf{x},t) + \mu(\mathbf{x},t) ) \nabla \cdot \mathbf{u}(\mathbf{x},t), \\[12pt]
\end{array}
\end{equation}
where $\mu$ is the artificial bulk viscosity.

These equations are discretized in time and space in 1D geometry as follows:
\begin{equation}
\label{eq:fluidUpdateDiscretized}
\begin{array}{{l}}
\rho_{m,g}^{n+1/2}  \displaystyle{\frac{u^{n+1}_g - u^n_g}  {\Delta t}} =  - \displaystyle{\frac{p^{n}_{g+1/2} + \mu^{n}_{g+1/2}  - p^n_{g-1/2} - \mu^{n}_{g-1/2}}  {\Delta x}}  \\  \quad \quad \quad \quad \quad \quad \quad \quad  \quad+ \rho_{c,g} ^{n+1/2}  E_{g}^{n}  \\[12pt]
\rho_{m,g+1/2}^{n+1/2}  \displaystyle{\frac{I^{n+1}_{g+1/2} - I^n_{g+1/2}} {\Delta t}}  =  - (p^n_{g+1/2} + \mu^n_{g+1/2}) \\ \quad \quad \quad \quad \quad \quad \quad \quad \quad \quad \quad \quad \displaystyle{\frac{u^{n+1/2}_{g+1} - u^{n+1/2}_{g}}  {\Delta x}}, \\[12pt]
\end{array}
\end{equation}
where $n$ is the time level of the discretization and $p^n = \rho_{m}^n I^n (\gamma - 1)$. 

In this work, we use an artificial bulk viscosity $\mu$ that has been proposed by \citet{Kuropatenko:1966,Chandrasekhar:1961}. This artificial bulk viscosity is non-zero only on grid cells for which $\nabla\cdot\mathbf{u}>0$ and is formulated as follows~\citep{Caramana:1998},
\begin{equation}
\label{eq:Kuropatenko}\mu= \rho_m \left[ c_2\frac{\gamma+1}{4} \left| \Delta\mathbf{u} \right| + \sqrt{c_2^2\left( \frac{\gamma+1}{4} \right)^2 \left( \Delta\mathbf{u} \right)^2 + c_1^2 c_s^2} \right] \left| \Delta\mathbf{u} \right|,
\end{equation}
where $\left|\Delta\mathbf{u}\right| = \left| \Delta u_x + \Delta u_y + \Delta u_z \right|$ is the velocity jump across the grid cell, $c_s = \sqrt{\gamma p / \rho_m}$ is the adiabatic sound speed, $c_1$ and $c_2$ are constants. 

{\bf \circled{4} Update Particle Quantities (Lagrangian Step).} In this phase, the new particle quantities are calculated to perform advection of the fluid and kinetic quantities. We use an explicit time-marching to advance both fluid and kinetic equations in time. 

Kinetic and fluid particle are updated in different ways as they advect different quantities in the fluid and kinetic PIC methods:
\begin{itemize}
\item Each {\bf fluid particle} quantity is updated by solving the Ordinary Differential Equations (ODEs):
\begin{equation}
\label{motionFluid}
\begin{array}{l}
 \displaystyle{\frac{d {\bf v}_p}{dt}} =   \displaystyle{\left. \frac{d \mathbf{u}}{dt} \right \vert_{\mathbf{x}_p} } \\[14pt]
  \displaystyle{\frac{d e_p}{dt}} =   m_p \displaystyle{\left. \frac{d I}{dt} \right\vert_{\mathbf{x}_p}} \\[14pt]
  \displaystyle{\frac{d {\bf x}_p}{dt}} = \mathbf{v}_p. 
\end{array}
\end{equation}
The discretized equations in 1D, using changes in fluid velocity and internal energy to reduce numerical diffusion~\citep{brackbill1986flip,brackbill1988flip}, are:
\begin{equation}
\begin{array}{l}
v_p ^ {n+1} = v_p ^ {n} +              (u^{n+1} - u^n) \vert_{x_p}\\
e_p ^ {n+1} = e_p ^ {n} +              m_p  (I^{n+1} - I^n)   \vert_{x_p }\\
x_p ^ {n+1} = x_p ^ {n} + v_p ^ {n + 1/2} \Delta t. 
\end{array}
\end{equation}
We note that  particle position is updated using the previous fluid particle velocity, differently from the fluid PIC method. Interpolated quantities at particle positions are calculated using interpolation functions (this time from particle onto grid):
\begin{equation}
\{u^{n+1} - u^n, I^{n+1} - I^n \}\vert_{\mathbf{x}_p} =\sum_g^{N_g} \{ u_g^{n+1} - u_g^{n}, I_g^{n+1} - I_g^{n}\} W( x_g - x_p^n). 
\end{equation}

\item Each {\bf kinetic particle} quantity is updated by solving the ODEs:
\begin{equation}
\label{motionKin}
\begin{array}{l}
 \displaystyle{\frac{d {\bf v}_p}{dt}} =  \frac{q}{m} {\bf E}\vert_{\mathbf{x}_p}\\[8pt]
 \displaystyle{\frac{d {\bf x}_p}{dt}} ={\bf v}_p.  
\end{array}
\end{equation}
The discretized equations are:
\begin{equation}
\begin{array}{l}
v_p ^ {n+1} = v_p ^ {n}  + \displaystyle{ \left. \frac{q}{m} E^n \vert_{x_p} \Delta t \right. } \\[8pt]
x_p ^ {n+1} = x_p ^ {n} + v_p ^ {n + 1/2} \Delta t. 
\end{array}
\end{equation}

The electric field at the particle position is calculated from the values of the electric field defined on the grid using the interpolation function as $E^n\vert_{x_p} =\sum_g^{N_g}  E_g^n W(x_g-  x_p^n)$.
\end{itemize}

{\bf \circled{5} Transforming a Fluid Particle to Kinetic Particle?} The PolyPIC algorithm allows us to dynamically flip a particle's type from fluid to kinetic, according to a predefined rule. It is possible to define several rules depending on the problem under study. An obvious choice is to switch from fluid to kinetic particles when fluid particles reach a threshold velocity or acceleration (in practice, we found that a multiple of local thermal velocity is a convenient threshold velocity):
\begin{equation}
\begin{array}{l}
\vert v_p^{n + 1} \vert >  c_1 v_{th} \quad  \textrm{or} \\[8pt]
\vert v_p^{n+1} - v_p^{n} \vert  >  c_2 v_{th}
\end{array}
\end{equation}

\begin{figure}[h!]
\begin{center}
\includegraphics[width=\columnwidth]{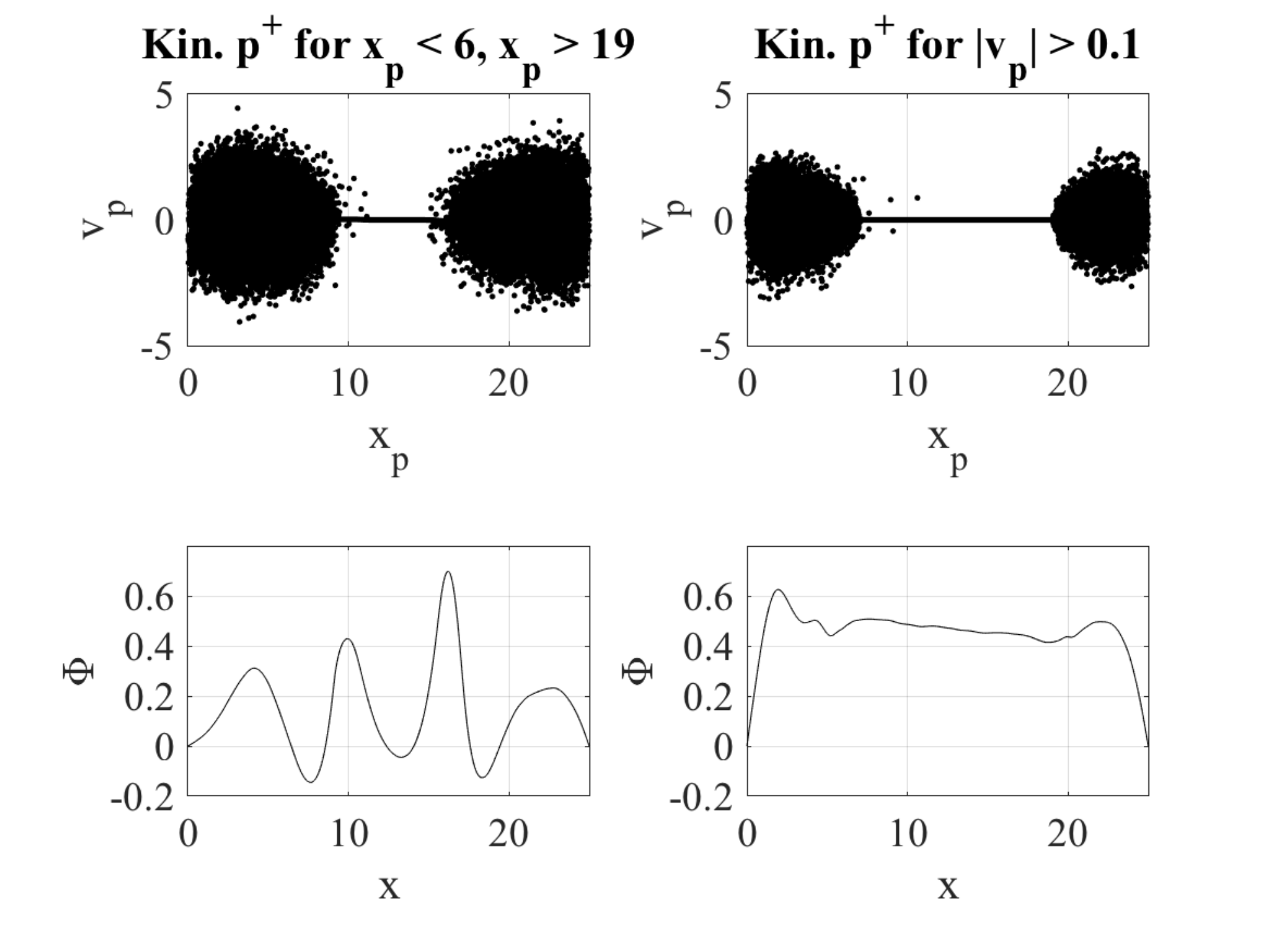}
\end{center}
\caption{Ion phase space and electrostatic potential $\Phi$ for a simulation with ions becoming kinetic when entering the regions $x < 6$ and $x > 15$ (left panels). An artificial sheath between the kinetic and the fluid ions is formed. In the right panels, ion phase space and electrostatic potential $\Phi$ is shown for a simulation with ions becoming kinetic when their velocity is greater than a threshold velocity ($0.1$). In this case, the artificial sheath is not formed.}\label{fig:Adaptivity}
\end{figure}

Similarly to other approaches in fluid-kinetic coupling, another obvious choice is to switch to kinetic particles in the regions where the kinetic effects are relevant. For instance, when studying plasma sheath formation close to a wall, it is useful to have kinetic electrons and/or ions close to the wall to model the sheath kinetically. This case is shown in the left panels of Figure \ref{fig:Adaptivity} where fluid ions become kinetic when entering the spatial regions $x < 6$ and  $x > 19$. However, we found that this choice creates numerical artifacts between the fluid and kinetic regions. Indeed, our experiments show that if we confine kinetic particles in a given spatial region, an artificial sheath forms at the interface between the kinetic and fluid regions. The formation of this artificial sheath is clear when analyzing the potential $\Phi$ profile in proximity of $x = 10$ and $x = 17$ (bottom left panel of Figure \ref{fig:Adaptivity}).
For this reason, in this work we do not switch to kinetic particles in selected parts of the domain. Instead, we choose a rule based either on reaching a threshold velocity or acceleration so that the transition is smoother and no evident sheath forms between fluid and kinetic regions (bottom right panel of Figure \ref{fig:Adaptivity}). 

When a fluid particle becomes kinetic, it obtains a new velocity. This velocity is sampled randomly from a Maxwellian distribution centered on the local fluid velocity $\left. \mathbf{u} \right\vert_{\mathbf{x}_p}$, with variance ($\sigma^2$) equal to the local thermal velocity: 
\begin{equation}
v_p = u \vert_{x_p} + \textrm{randn}(\sigma^2 = v_{th})
\end{equation}

The local (to the particle) fluid and thermal velocity ($v_{th}$) are calculated using interpolation functions as
\begin{equation}
\begin{array}{l}
u \vert_{x_p}  = \sum_g^{N_g} u_g W( x_g- x_p)   \\[8pt]
v_{th} =\sum_g^{N_g} \sqrt{p_g/ \rho_{m,g}} W( x_g- x_p). 
\end{array}
\end{equation}

In the transformation from fluid to kinetic particle, we assume a Maxwellian distribution function for kinetic particles while different kinds of distribution function might occur in non-equilibrium plasmas.

\subsubsection{Numerical Stability}
Both fluid and kinetic PIC methods in this work use an explicit discretization in time, and are subject to their respective numerical stability constraints:
\begin{itemize}
\item Because we use an explicit formulation of equations in \circled{3}, the fluid PIC simulation time step and grid spacing must satisfy the Courant condition $\Delta t \leq \Delta x / c_s = \Delta x / \sqrt{\gamma (\gamma-1) I}$. An implicit discretization of equations with pressure term evaluated half time (stage \circled{3}) would remove this stability condition~\citep{brackbill1986flip,brackbill1988flip}. In addition, explicit fluid PIC methods are unstable against the ringing instability when the plasma flow is lower than a critical velocity~\citep{brackbill1988ringing}.

\item The kinetic PIC component requires a time step resolving the plasma period $\omega_p \Delta t \leq 0.1$ to retain numerical stability. In addition, grid spacing should be smaller than the local Debye length ($\Lambda_D$) to avoid numerical heating and the finite grid instability (see, e.g., \citet{birdsall2004plasma}).
\end{itemize}


\subsubsection{Implementation}
We implement the PolyPIC method in a proof-of-concept Matlab code, available at \href{http://www.github.com/smarkidis/}{http://www.github.com/smarkidis/}. In our implementation, we use only vector operations with masks to avoid conditional branching and achieve increased performance. The Poisson equation requires the solution of a linear system that is calculated with the Matlab solver for tridiagonal matrices. The interpolation operations that are implemented as a large sparse matrix vector multiplications take most of the simulation time. In most of the simulations presented in Section~\ref{results}, interpolation operations account for more than 50\% of the total simulation time.

The interpolation step in phase \circled{1} of the PolyPIC method requires interpolation of both fluid and kinetic particles onto the grid. Because of the thermal spread of kinetic particles, the fluid quantities calculated with kinetic particles are affected by numerical noise. This noise appears as relatively small discontinuities in the fluid quantities, densities, fluid velocity and pressure. During the update of the fluid quantities in step \circled{3}, spurious oscillations might originate because of these small-scale discontinuities. 

To address this problem, we use artificial bulk viscosity in Equation \ref{eq:fluidUpdateDiscretized} to dissipate these spurious oscillations. In addition, a Laplacian smoothing of fluid quantities is beneficial to eliminate small-scale discontinuities in fluid quantities~\citep{birdsall2004plasma,stanier2018fully} before solving the fluid equations on the grid (phase \circled{3}). The Laplacian smoothing operation in one dimension on a grid quantity $Q$ is defined as follows:
\begin{equation}
\label{eq:smoothing}
S(Q_g) = \frac{Q_{g-1} - 2Q_g + Q_{g+1}}{4}.
\end{equation}
More than one smoothing pass can be also performed as $S(...S(Q)...)$.



\section{Results}
\label{results}
We present four different verification tests of the PolyPIC model. The first two tests are the \emph{Landau damping of the Langmuir} and \emph{ion acoustic waves} tests, showing the use of fluid ions and kinetic electrons (Langmuir wave test) and kinetic ions and fluid electrons (ion acoustic wave test). The third and fourth tests, the \emph{two-stream instability} and \emph{sheath formation} tests, show the dynamic change from fluid to kinetic description within one simulation. All the tests are performed in one-dimensional geometry in the electrostatic limit.

\subsection{Landau Damping of Langmuir Waves}
The first test is the simulation of a Langmuir wave propagation in a plasma. A Langmuir wave undergoes Landau damping due to kinetic resonance between the wave and electrons moving approximately at the phase velocity of the Langmuir wave. The kinetic energy of such electron population increases at expense of the Langmuir wave damping. For this reason, a kinetic treatment of electrons is required for modeling the Landau damping of Langmuir waves.

We perform a simulation of the Langmuir wave propagation using one population of kinetic electron particles and one population of fluid ion particles. The initial electron thermal velocity is $V_{the} = 1$ with equal temperature for electrons and ions. The charge to mass ratio is set to $-1$ for electrons and to $1/1836$ for ions. There are $10,000$ kinetic electrons and fluid ions per cell. The simulation box of size $L = 4 \pi \Lambda_D$ is divided in $64$ cells and has periodic boundaries. To initiate the Langmuir wave we perturb the initial positions of kinetic electrons: $x_p = x_p + 0.1\sin(2 \pi x_p / L)$. The simulation step is $\Delta t = 0.1/\omega_p$; a simulation lasts for $150$ computational cycles. The specific heat ratio for the fluid ions is $\gamma=7/5$. We perform a two-pass binomial smoothing of the ion fluid quantities at each time step, and no artificial viscosity is introduced ($c_1=0$ and $c_2=0$ in Equation~\ref{eq:Kuropatenko}). 

\begin{figure}[h!]
\begin{center}
\includegraphics[width=\columnwidth]{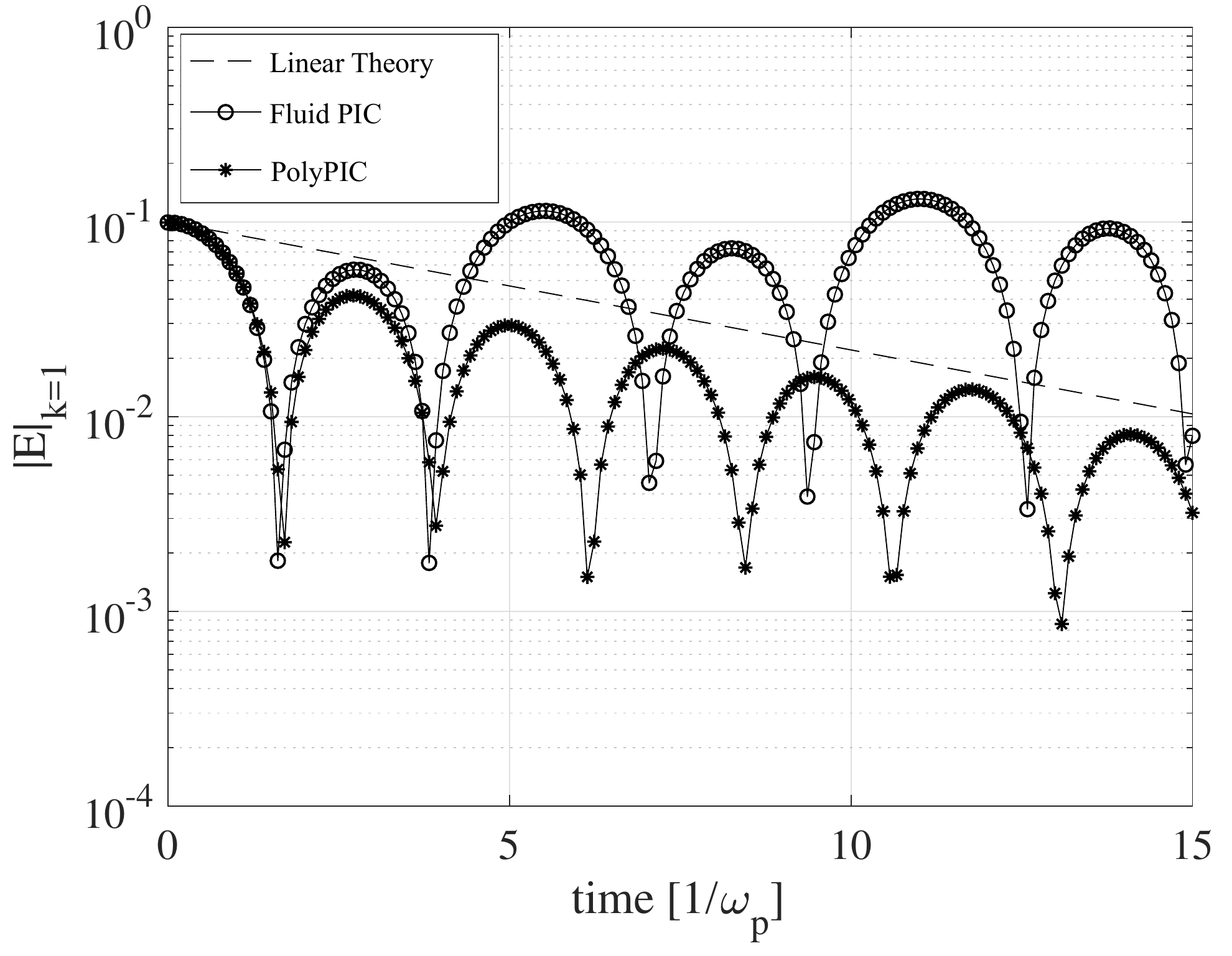}
\end{center}
\caption{Comparison between linear theory and PolyPIC simulation of Langmuir wave damping. The $k = 1/\Lambda_D$ spectral component of the electric field for the simulation with kinetic electrons and fluid ions (asterisks) decreases as predicted by the linear theory (dashed line). The fluid PIC simulation with both fluid electron and ions (open circles) do not show damping of the Langmuir wave.}\label{Langmuir}
\end{figure}

The $k = 1/\Lambda_D$ spectral component of the electric field in Figure~\ref{Langmuir} (asterisks) is compared with the damping rate obtained from the linear theory $\gamma = -0.15139 \omega_p$ (dashed line) and simulation results of a fluid PIC.
To assess the importance of kinetic electrons, we also perform a simulation of Langmuir wave propagation with a fluid PIC simulation. The open circles in Figure~\ref{Langmuir} represent the electric field spectral component for the fluid PIC simulation, showing that the Langmuir wave is not damped when fluid approach is used.

\subsection{Landau Damping of Ion Acoustic Waves}
The second test for the PolyPIC method is similar in nature to the first test of Langmuir wave propagation as it investigates the kinetic damping of an ion acoustic wave. Differently from the previous test, we use fluid electrons and kinetic ions in PolyPIC. As in the previous test, the ion acoustic wave is expected to damp in time because of kinetic effects.
The simulation is initialized with $10,000$ fluid electrons and kinetic ions per cell. The charge/mass ratio is $-1$ for electrons and $1/1836$ for ions. The periodic simulation box of size $L = 4 \pi$ is divided in $64$ cells. The ion acoustic wave is excited by perturbing the initial positions of the kinetic ions: $x_p = x_p + 0.2 \sin(2 \pi x_p / L)$. The initial thermal velocity of ions is $V_{thi} = \sqrt{1/3}$ and electron/ion temperature ratio is $T_e/T_i=5$. A simulation lasts for $600$ computational cycles with $\Delta t = 0.025/\omega_p$. The specific heats ratio for fluid ions is $\gamma=5/3$. We perform a two-pass Laplacian smoothing (Equation~\ref{eq:smoothing}) of the ion fluid quantities at each time step before phase \circled{3}, and artificial viscosity with $c_1 = 10$ and $c_2 = 10$ is used in Equation \ref{eq:fluidUpdateDiscretized}. 

\begin{figure}[h!]
\begin{center}
\includegraphics[width=\columnwidth]{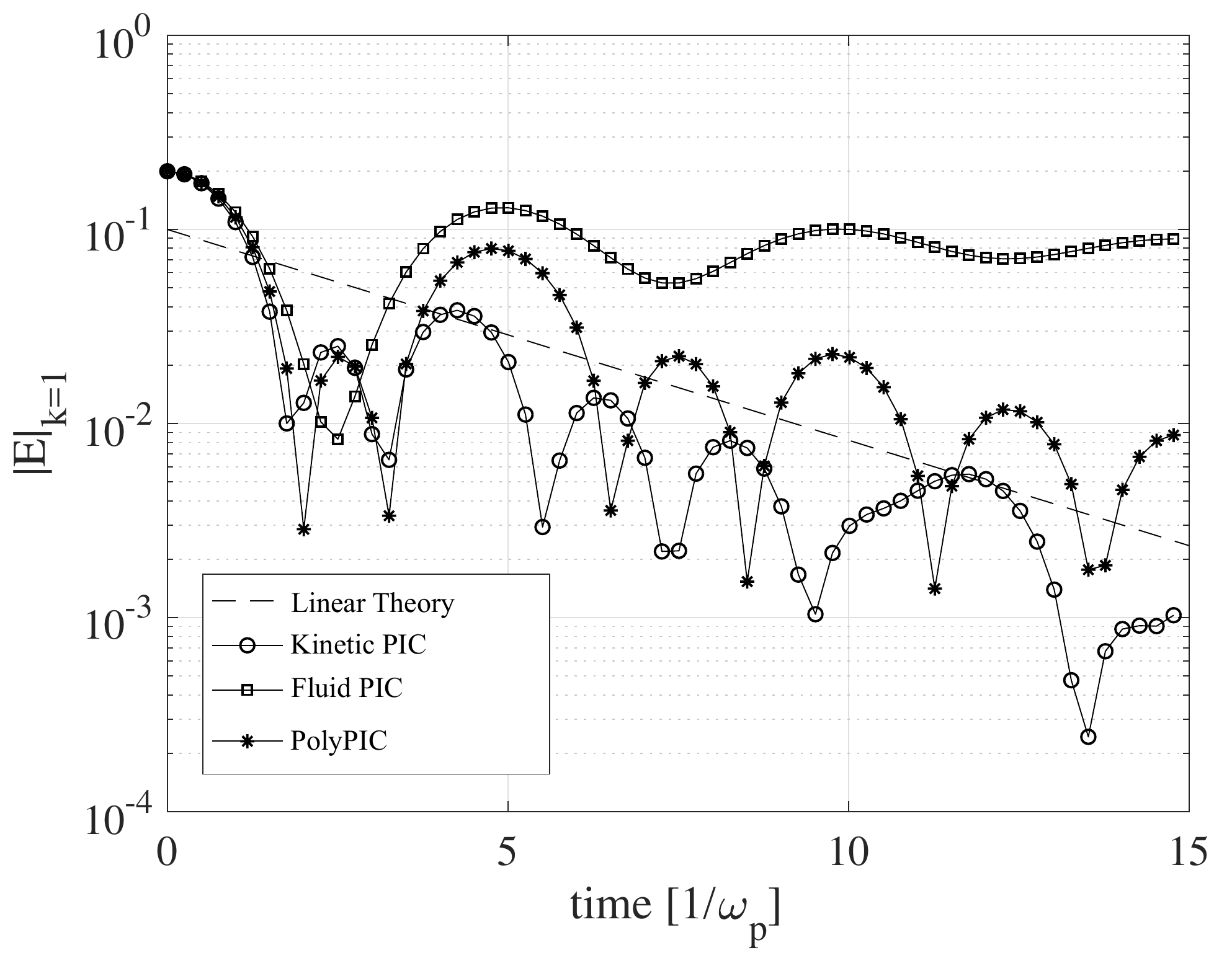}
\end{center}
\caption{Simulation of ion acoustic propagation in plasma using fully kinetic simulation (open circles), fluid electrons and kinetic ions (asterisks) and fully fluid (open squares) simulations.}\label{IonAcoustic}
\end{figure}

The ion acoustic wave is damped by the resonant interaction of the wave with particles as shown in Figure~\ref{IonAcoustic}, where asterisks depict the electric field's first spectral component in the simulation with PolyPIC. A theoretical damping rate $-0.25 \omega_p$ is plotted with dashed line for comparison.
In addition, the results of kinetic PIC with $10,000$ electrons and ions per cell  (open circles) and fluid PIC (open squares) simulations are shown in Figure~\ref{IonAcoustic}. The fully kinetic simulation shows a damping rate similar to the rate in the PolyPIC simulation. However, the wave trapping period differs and at the end of the simulation the effects of numerical noise become evident. Additional kinetic PIC simulations with larger number of particles shows a decrease of numerical noise and a convergence to correct representation of wave-particle interaction. On the other hand, fluid PIC simulation misrepresents the physics, and the ion acoustic wave is not damped.

\subsection{Two-Stream Instability}
The \emph{two-stream instability} test aims at verifying the dynamic change from fluid to kinetic electrons. The two-stream instability is a kinetic instability occurring in presence of two counter-streaming electron beams. The linear growth of the instability can be described correctly by two electron fluid theory~\citep{birdsall2004plasma}. For this reason, the PolyPIC simulation can start from fluid electrons where each beam constitutes a fluid. However, the non-linear part of the instability cannot be described correctly by the two fluid theory and a kinetic treatment is required. During the non-linear part of the instability, many of the fluid electrons undergo strong acceleration and flip their type from fluid to kinetic.

The simulation initializes the two electron beams as two separate fluid particle species with 100 particles/beam/cell in a periodic domain with $L = 2 \pi / 3.06$ divided into 64 grid cells. We initialize the electron beams of relatively \emph{cold} electrons with thermal speed $V_{the} = 0.01$, and beam drift velocity $\pm 0.2$. Motionless ions provide only a background charge density to neutralize the system (they are not explicitly considered in the simulation). The simulation lasts for $2,100$ cycles with $\Delta t = 0.02/\omega_p$.  The time step is chosen to resolve electron dynamics during instability and satisfies the numerical stability constraints. The two-stream instability is initiated by perturbing the initial electron positions: $x_p = x_p + 0.1\sin(2 \pi x_p / L)$. The specific heats ratio for electrons in this simulation is $\gamma=7/5$. An artificial viscosity (Equation~\ref{eq:Kuropatenko}) is used with $c_1 = 1$ and $c_2 = 1$.

In the PolyPIC simulation, fluid electrons become kinetic when they undergo a strong acceleration. Namely, when a particle's fluid velocity variation in a time step $v_p^{n+1} - v_p^{n}$ is larger than $V_{the}/10$. We found that in practice such a threshold value allows fluid particles to become kinetic during the initial stage of the non-linear part of the instability.

\begin{figure}[h!]
\begin{center}
\includegraphics[width=\columnwidth]{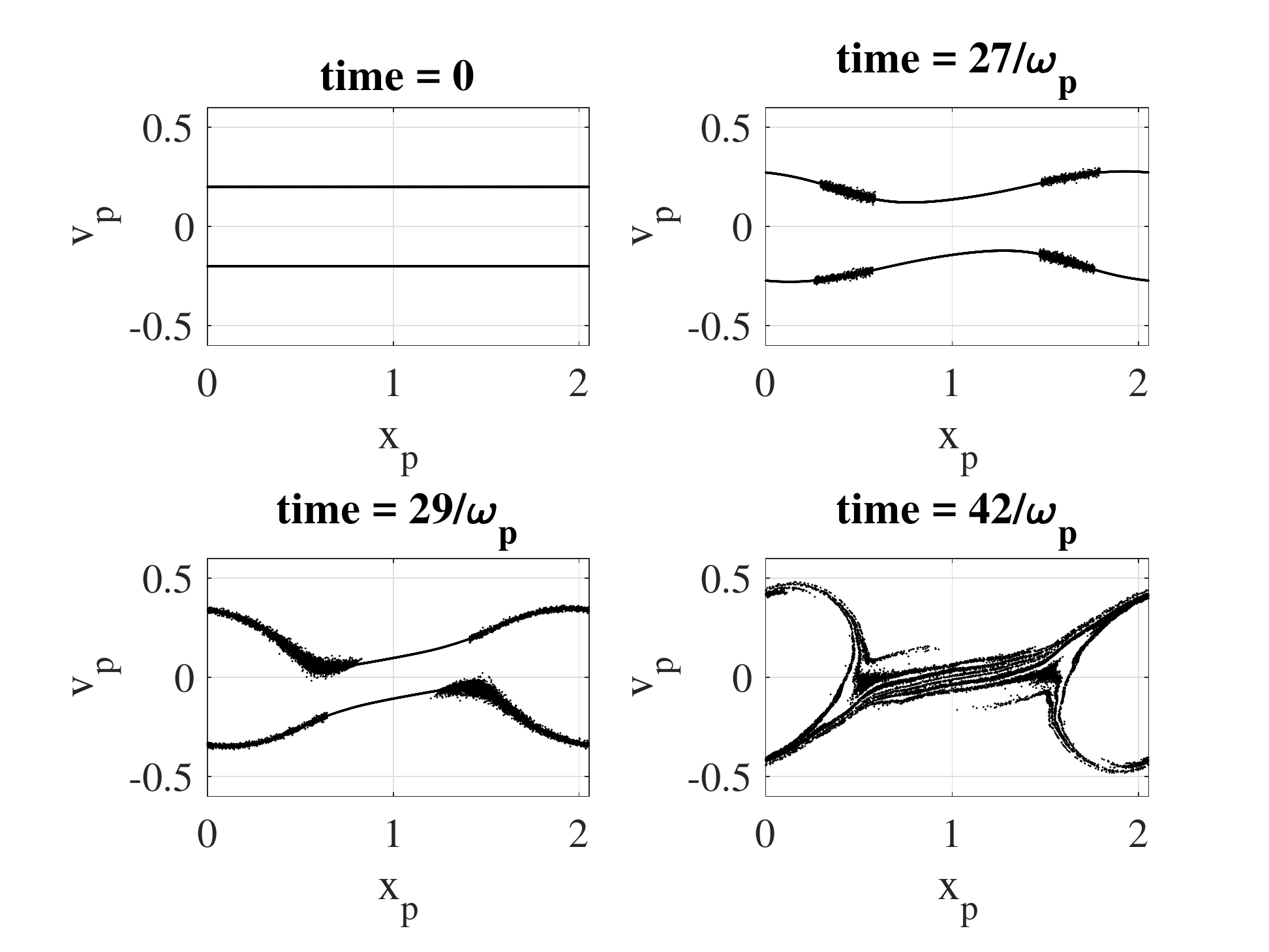}
\end{center}
\caption{Electron phase space during the two-stream instability at different times. Initially all the electrons are fluid. Electrons undergoing strong acceleration become kinetic. This clear from inspecting the electron phase-space at $time = 27, 29 / \omega_p$. At $time = 42 / \omega_p$ all the electrons are kinetic. A video of electron phase space is available \href{https://github.com/smarkidis/fluid-kinetic-PIC/blob/master/TwoStream.mov}{here}.}\label{fig:twostream}
\end{figure}

Initially in the PolyPIC simulation, the two electron beams consist of only fluid particles. This is clear from inspecting the top left panel of Figure \ref{fig:twostream} showing the electron phase space at time 0. Approximately at $t = 27/\omega_p$, fluid electrons undergoing a strong acceleration start turning to kinetic. This is visible by the spread of electrons due to thermal noise in the four regions in the top right panel of Figure~\ref{fig:twostream}. The extent of these \emph{kinetic} regions increases with time until a certain moment when it covers the entire simulation domain. Finally, all electrons are kinetic at time $t = 42 / \omega_p$ (bottom right panel of Figure \ref{fig:twostream}).

\begin{figure}[h!]
\begin{center}
\includegraphics[width=\columnwidth]{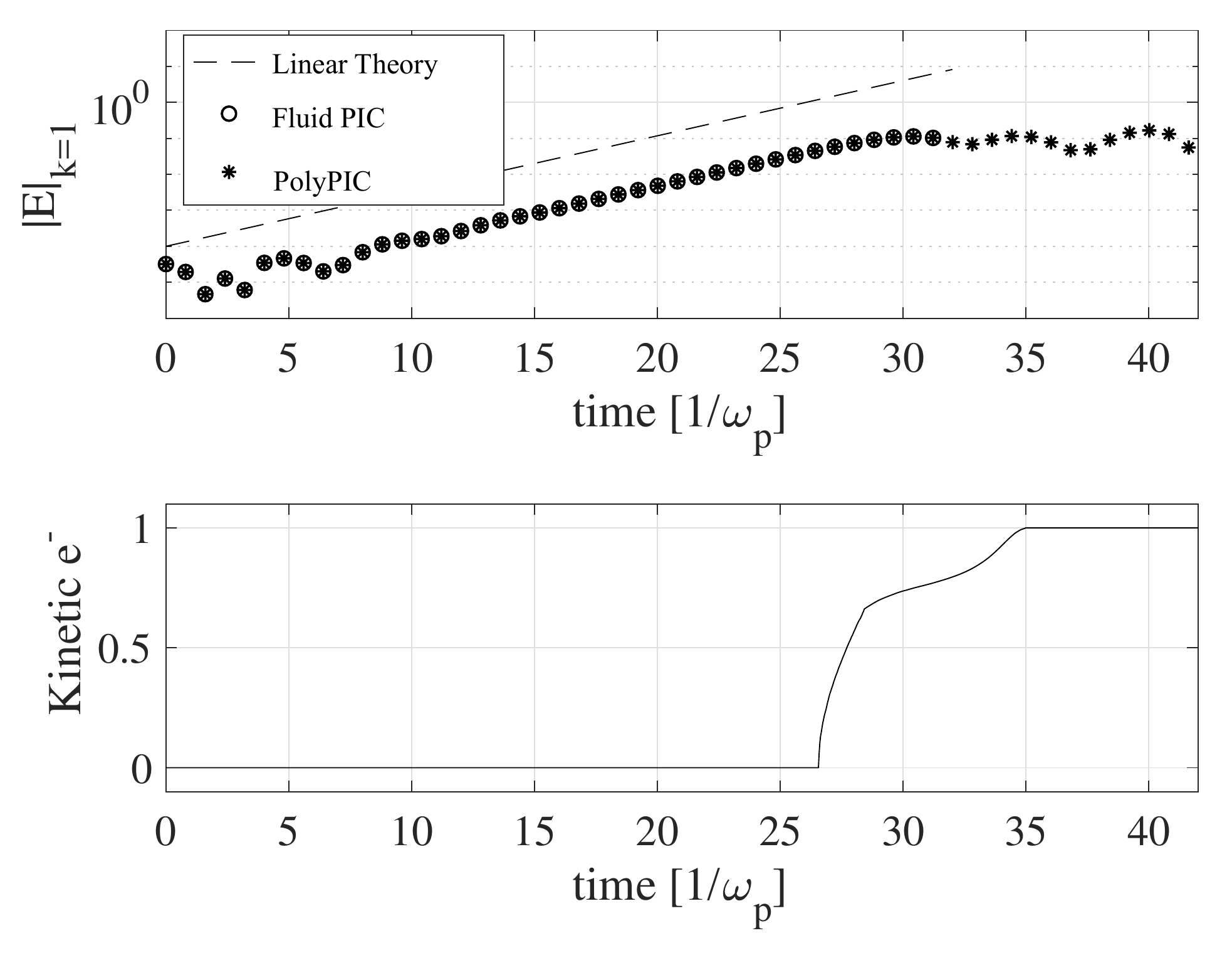}
\end{center}
\caption{The $k = 1$ spectral component of the electric field is shown for two fluid PIC simulation (open circles), PolyPIC simulation (asterisks) and linear theory (dashed line) in the top panel. The two fluid PIC simulation becomes unstable during the non-linear stage of the instability. The bottom panel shows the ratio of kinetic electrons over the total number of electrons during the PolyPIC simulation. Initially, all the electrons are fluids; during the non-linear phase of the instability, the electrons become kinetic.}\label{fig:twostreamLT}
\end{figure}

The  PolyPIC method is verified against linear theory prediction of the instability growth rate. The top panel of Figure~\ref{fig:twostreamLT} shows a comparison of the simulated electric field component $k = 1$ (asterisks) with the instability growth rate of $0.35355 \omega_p$, predicted by the linear theory (dashed line). Open circles show the growth of the instability in the two fluid PIC simulation. The two fluid PIC simulation models correctly the linear stage of the instability growth, but then becomes unstable at $t = 31/\omega_p$. We note that higher values of artificial viscosity allows the simulation to progress for longer period in the non-linear regime of the instability, but eventually the fluid simulation becomes numerically unstable.
The bottom panel of Figure~\ref{fig:twostreamLT} shows the ratio of kinetic electrons over the total number of electrons in the PolyPIC simulation. Initially, all the electrons are fluid. The first electrons become kinetic approximately at $t = 26/\omega_p$. The electrons are all kinetic at $t = 31/\omega_p$.

\subsection{Plasma Sheaths}
The last problem used to test the PolyPIC method is the sheath formation in the proximity of walls. Because of the higher mobility of electrons, initially more electrons exit than ions, leading plasma to have positive potential with respect to the wall. 

This simulation is initiated with kinetic electrons and fluid ions. During the simulation, ions can become kinetic if their velocity is greater than a specific threshold velocity. Ions are accelerated close to the domain walls, hence we expect the kinetic regions to form adjacent to the walls. The simulation box with $L=25 \Lambda_D$ long is divided into $256$ cells. We eliminate particles exiting the simulation box and fix the electrostatic potential on the walls to $0$. Initially, there are $500$ electrons and ions per cell. The charge/mass ratio is $-1$ for electrons and $1/1836$ for ions. The initial thermal velocity of electrons is $V_{the} = 1$ and electron/ion temperature ratio is $1$. The fluid ions become kinetic when they reach a threshold velocity of $40V_{the}$. The simulation lasts for $2,000$ computational cycles with $\Delta t = 0.05/\omega_p$. The specific heats ratio for the fluid ions is $\gamma=7/5$. We perform a two-pass binomial smoothing of the ion fluid quantities, and an artificial viscosity (Equation~\ref{eq:Kuropatenko}) is used with $c_1 = 1$ and $c_2 = 1$. 

\begin{figure}[h!]
\begin{center}
\includegraphics[width=\columnwidth]{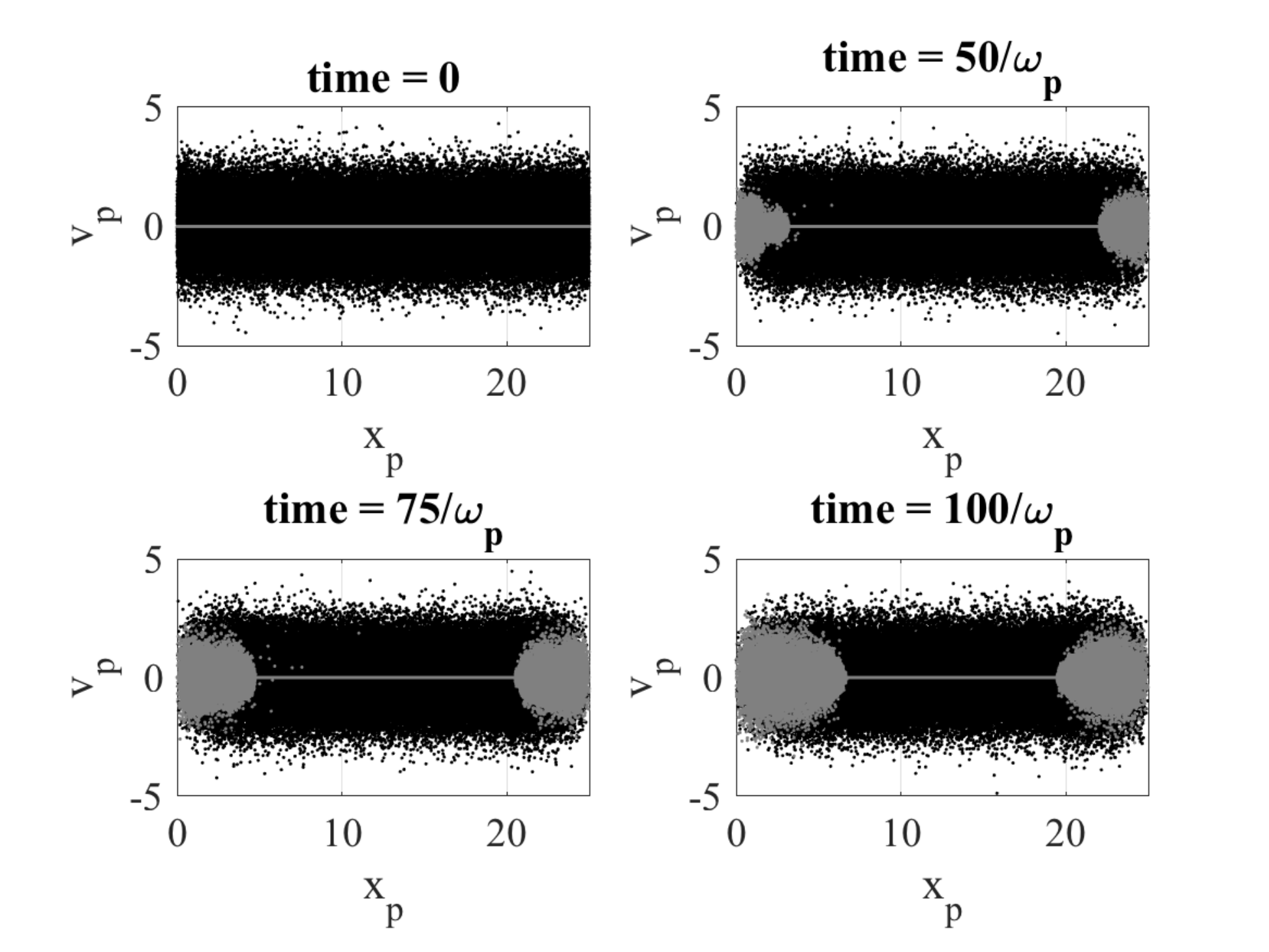}
\end{center}
\caption{Electron (black dots) and ion (grey dots) phase space in the \emph{sheath} simulation. Initially all the electrons are kinetic and ions are fluid. If ion particle velocity becomes higher than 40 times the initial ion thermal velocity, the ion particle becomes kinetic. Fluid ion particles become kinetic in the sheath close to the walls ($x=0$ and $x=L$). A video of phase space is available \href{https://github.com/smarkidis/fluid-kinetic-PIC/blob/master/Sheath.mov}{here}.}\label{WallCharging}
\end{figure}

\begin{figure}[h!]
\begin{center}
\includegraphics[width=\columnwidth]{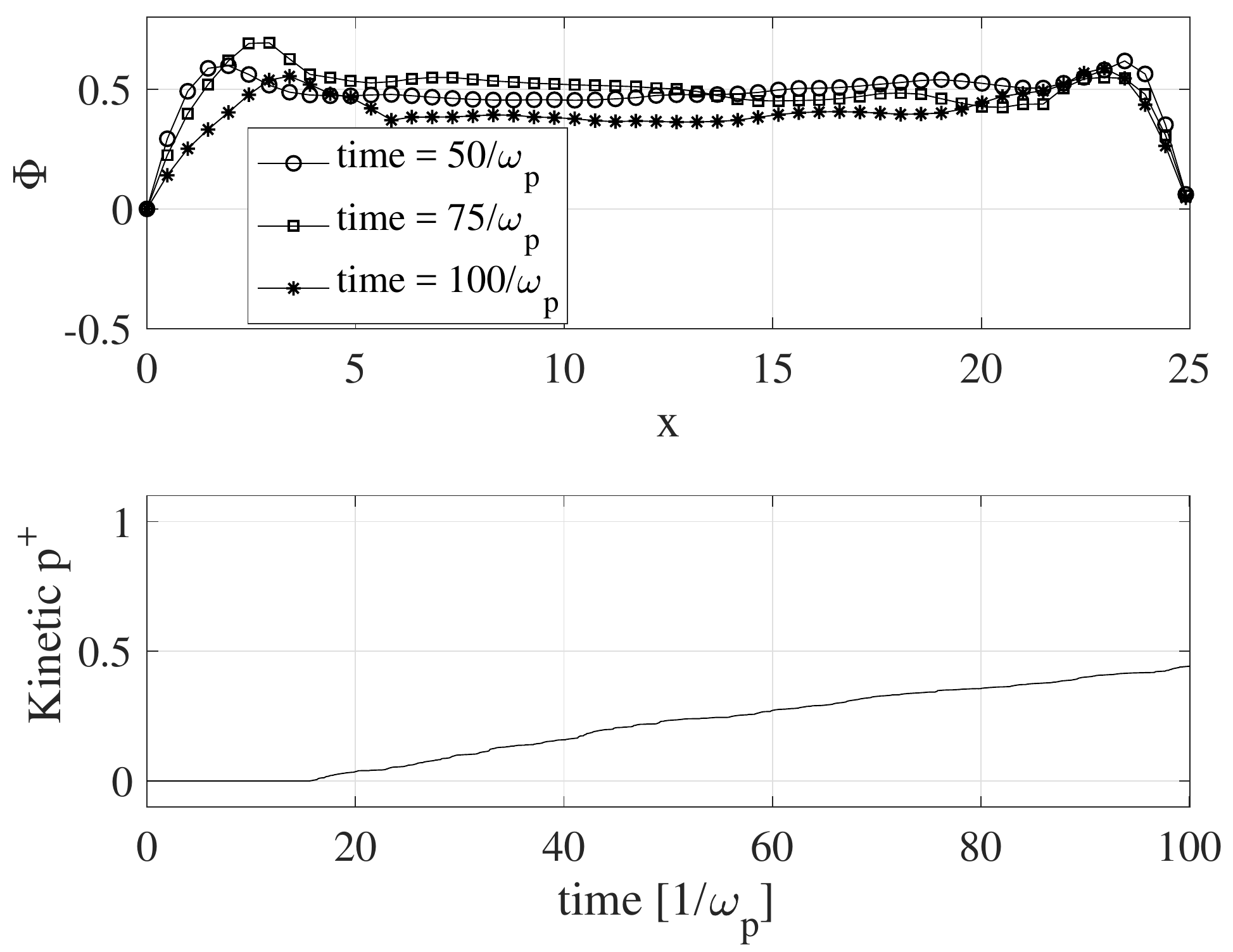}
\end{center}
\caption{Top panel: the electrostatic potential in the simulation domain at three time steps: $t\omega_p=50$ (open circles), $t=75/\omega_p$ (open squares), $t=100/\omega_p$ (asterisks). Bottom panel: the ratio of kinetic ions during the \emph{sheath} simulation.}\label{WallChargingPer}
\end{figure}

Initially all ions in the simulation are fluid and depicted with grey dots in the phase space plot shown in the upper left panel of Figure~\ref{WallCharging}. As the simulation progresses, in proximity of the walls ions reach velocity higher than the threshold velocity and become kinetic. Subsequent panels in Figure~\ref{WallCharging} show the widening regions adjacent to the walls where kinetic ions are spread by thermal noise.
The top panel of Figure~\ref{WallChargingPer} shows electrostatic potential at different simulation times. The electrostatic potential remains similar during the simulation and it is not impacted by ions switching from fluid to kinetic. The bottom panel of Figure~\ref{WallChargingPer} shows the ratio of kinetic ions over the total number of ions in the PolyPIC simulation. Initially all the ions are fluid. At $t = 18/\omega_p$ the first ions start turning to kinetic; after this, the number of kinetic ions increases linearly with time.

\section{Discussion and Conclusion}
\label{conclusion}
A new PIC method using polymorphic computational particles has been formulated and implemented to combine the fluid and kinetic PIC methods under a unified common framework. The PolyPIC method is adaptive as it allows fluid particles to become kinetic when undergoing a strong acceleration or reaching a threshold velocity. We implemented a proof-of-concept code based on the explicit discretization of the governing equations and used it to solve successfully four different test problems.

A first challenge when coupling fluid and kinetic approaches is to eliminate spurious effects occurring at the boundary between fluid and kinetic regions. For instance, we showed that a sheath forms at the interface between the regions with kinetic particles and with fluid particles in the same way a sheath forms when two different plasmas are put in contact. It is preferable to change the type of the particle from fluid to kinetic depending on its velocity instead of its position to allow a smooth transition between the fluid and kinetic regions. Currently, the criterion to switch particles from fluid to kinetic are set empirically. A dedicated study is needed on how to set these criteria automatically.

A second major challenge in coupling fluid and kinetic approaches in the same PIC method is that moments (densities, fluid velocity, ...) computed from kinetic particles are not smooth, as fluid quantities typically are, because of the kinetic particle noise. This noise produces small discontinuities in the computed moments which might result in unphysical oscillations. An artificial bulk viscosity introduced into fluid equations effectively remedies the effects of numerical noise. In addition, smoothing and filtering can reduce noise from kinetic particles before updating the fluid quantities. How to eliminate such spurious effects without affecting energy conservation, is a topic of future research.

One of the advantages of fluid PIC method with respect to the kinetic PIC approach is the fact fluid PIC method requires only a small number of particles to describe accurately the evolution of the system. However, kinetic PIC methods typically require a very large number of particles to describe accurately kinetic effects such as wave-particle interaction. In order to use few fluid particles but many kinetic particles when needed, the PolyPIC method requires a particle splitting technique when switching from one fluid particle to many kinetic particles~\citep{lapenta1994dynamic}.

In this work we did not address the change of a kinetic particle to fluid particle. Presently there is no clear and simple use case to present. However, such techniques can be easily designed. An efficient implementation might require the coalescence of several kinetic particles in one fluid particle.

To conclude, we have unified the fluid and kinetic PIC methods under a unified common framework comprising both fluid and kinetic particles. This approach allows the simulation to change smoothly from fluid to kinetic description in time and space providing a powerful tool for adaptive fluid-kinetic plasma simulations.

\end{document}